\documentclass[aps,prd,showpacs,nobibnotes,twocolumn,
nobalancelastpage,superscriptaddress]{revtex4}
\usepackage{graphicx}
\usepackage{amsmath}
\def\gap{\;\rlap{\lower 2.5pt
 \hbox{$\sim$}}\raise 1.5pt\hbox{$>$}\;}
\def\lap{\;\rlap{\lower 2.5pt
   \hbox{$\sim$}}\raise 1.5pt\hbox{$<$}\;}
\def\gsim{\;\rlap{\lower 2.5pt
 \hbox{$\sim$}}\raise 1.5pt\hbox{$>$}\;}
\def\lsim{\;\rlap{\lower 2.5pt
   \hbox{$\sim$}}\raise 1.5pt\hbox{$<$}\;}
\def\msun{{\rm\,M_\odot}}

\begin{document}


\title{A New Signature of Dark Matter Annihilations:\\
Gamma-Rays from Intermediate-Mass Black Holes
}

\author{Gianfranco Bertone}
\affiliation{Particle Astrophysics Center, Fermi National
Accelerator Laboratory, Batavia, Illinois 60510-0500, USA}
\author{Andrew R. Zentner}
\affiliation{Kavli Institute for Cosmological Physics and 
Department of Astronomy and Astrophysics, 
The University of Chicago, Chicago, IL 60637,USA}
\author{Joseph Silk}
\affiliation{Astrophysics, Denys Wilkinson Building, 
Keble Road, Oxford OX1 3RH, United Kingdom}

\date{\today}

\begin{abstract}
We study the prospects for detecting gamma-rays 
from Dark Matter (DM) annihilations in 
enhancements of the DM density (mini-spikes) 
around intermediate-mass black holes with masses 
in the range $10^2 \lsim M / \msun \lsim 10^6$.
Focusing on two different IMBH formation scenarios, we show that, 
for typical values of  mass and cross section of common 
DM candidates, mini-spikes, produced by the adiabatic growth of
DM around pregalactic IMBHs, would be bright sources of gamma-rays, which 
could be easily detected with large field-of-view gamma-ray 
experiments such as GLAST, and further studied with 
smaller field-of-view, larger-area experiments like 
Air Cherenkov Telescopes CANGAROO, HESS, MAGIC and VERITAS.  
The detection of 
many gamma-ray sources not associated with a luminous component of 
the Local Group, and with identical cut-offs in their energy spectra 
at the mass of the DM particle, would provide a potential smoking-gun 
signature of DM annihilations and shed new light on the nature 
of intermediate and supermassive Black Holes.  
\end{abstract}


\pacs{95.35.+d,97.60.Lf,98.62.Js,98.70.Sa,98.70.Rz
\hspace{2.2cm} FERMILAB-PUB-05-408-A}

\maketitle


\section{Introduction}

Although many astrophysical and cosmological observations provide 
convincing evidence for the existence of a 
``dark'' component in the matter density of the 
Universe,  the nature of this {\it dark matter} (DM) remains unkown.  
It is commonly assumed that DM is made of 
new, as yet undiscovered, particles, 
associated with theories beyond the 
Standard Model of Particle Physics.  
Among the most widely studied candidates 
are the supersymmetric neutralino and 
candidates arising in theories with extra-dimensions, which 
appear difficult to constrain with direct searches (i.e. by looking for 
nuclear recoils due to DM particles scattering off nuclei) and whose prospects
of discovery at future accelerators strongly depend on the details of the 
particle physics setup (for recent reviews see e.g. 
Refs.~\cite{Bergstrom:2000pn,Bertone:2004pz}).  
Indirect searches via the detection of annihilation radiation may 
provide an interesting alternative, but they are usually affected
by large astrophysical and cosmological uncertainties.  
Furthermore, in many cases, the detection of an annihilation 
signal may be difficult to distinguish from less exotic 
astrophysical sources. An example of this is the case of the 
Galactic center, where such high-energy radiation has been 
recently observed by several different experiments, 
without providing any conclusive evidence for or against 
an interpretation in terms of DM annihilation products 
(see Refs.~\cite{Aharonian:2004wa,Bergstrom:2004cy,Hooper:2004fh,
Bertone:2005hw,Bertone:2005xv} and references therein).

Here we describe a scenario that may provide smoking-gun 
evidence for the annihilation of DM particles.  
If intermediate-mass black holes (IMBHs), 
with a mass ranging between $10^2$ and $10^6 \msun$ 
(e.g.~\cite{Miller:2003sc}), exist in the Galaxy, 
their adiabatic growth would have
modified the DM distribution around them, leading  to the formation
of ``mini-spikes'', that is, large, local enhancements of the DM density
\cite{Zhao:2005zr}.  
The DM annihilation rate being proportional 
to the square of the number density of DM particles, 
these mini-spikes would be bright gamma-ray sources, 
distributed in a roughly spherically-symmetric way about 
the galactic center, and well within the observational 
reach of the next-generation gamma-ray experiments.  
Their brightness and isotropy make them ideal targets of large field-of-view 
gamma-ray experiments such as GLAST~\cite{glast}.  
In case of a positive detection, 
Air Cherenkov Telescopes such as 
CANGAROO~\cite{canga}, HESS~\cite{hess}, MAGIC~\cite{magic} and 
VERITAS~\cite{veritas} could extend 
the observations to higher energies and improve the angular resolution.
We argue that the observation of numerous (up to $\sim 100$) 
point-like gamma-ray sources with identical 
cut-offs in their energy spectra, at an energy equal to
the mass of the DM particle, would provide 
smoking-gun evidence for DM particles. 

In this paper, we make predictions for the number of 
detectable black holes in two different 
IMBH formation scenarios.  In the first scenario,
IMBHs form in rare, 
overdense regions at high redshift, 
$z \sim 20$, as remnants of Population III stars, and have a 
characteristic mass-scale of a few $10^2 \msun$ 
\cite{Madau:2001} (a similar scenario
was investigated in Ref.~\cite{Zhao:2005zr,islamc:2004,islamb:2004}).  
In this scenario, these black holes serve as the 
seeds for the growth supermassive black 
holes found in galactic spheriods \cite{Ferrarese:2005}.  
In the second scenario, IMBHs form directly out of cold gas 
in early-forming halos, in a sense that will be specified below, 
and and are typified by a larger mass scale, 
of order $10^5 \msun$. We demonstrate that, with respect to  
Ref.~\cite{Zhao:2005zr}, 
the latter scenario leads to qualitative differences in the 
mini-spike profiles with dramatic consequences for the 
detectability of gamma-ray fluxes. For both scenarios, we  make 
detailed estimates of the population of IMBH in the Milky Way DM halo 
using a complete model of IMBH formation at high redshift, 
black hole mergers, and halo merger and evolution 
\cite{Koushiappas:2005qz}.
This allows us the unique ability to make a detailed study 
of the detectabilty of mini-spikes as gamma-ray sources. 

This paper is organised as follows.  In Sec.~II~A, 
we review the evidence and formation scenarios 
for IMBHs.  In Sec.~II~B, we describe the 
model that we employ to estimate the properties of the local 
IMBH population, and we present the main properties 
(radial profile, mass function etc.) of IMBH 
populations in Milky Way-like halos in Sec.~II~C.  
Sec.~III is devoted to the calculation of the mini-spike profiles 
and Sec.~IV to the DM annihilation fluxes.  
Sec.~V contains our primary results on the 
observability of gamma-rays from the annihilation of 
DM around IMBHs.  In Sec.~VI, we discuss the 
implications of our results and draw our conclusions.
We perform all of our calculations in the context of a 
standard, flat cosmological constant plut cold 
DM ({$\Lambda$}CDM) cosmology with $\Omega_{\rm M}=0.3$, 
$\Omega_{\rm \Lambda}=0.7$, $h=0.7$ and a scale-invariant primordial 
power spectrum with a normalization set by $\sigma_8=0.9$.

\section{Evidence and Properties of IMBHs}

\subsection{The case for IMBHs}

%
%
In the last few years, observational and theoretical evidence has
accumulated ~\cite{Miller:2003sc} for the existence of compact 
objects, heavier than {\it stellar} black holes, but lighter than  
the so-called {\it supermassive} black holes (SMBHs) 
lying at the centers of galactic spheroids.  
We consider here the mass range 
$20 \lsim M_{\rm IMBH}/\msun \lsim 10^6$, 
where the lower bound of the IMBHs mass corresponds to 
recent estimates of the maximum mass of the remnant of a massive 
stellar collapse~\cite{Fryer:2001}, 
and the upper limit roughly indicates the minimum mass of SMBHs,
assumed to lie in the range $10^6 \lsim M_{\rm SMBH}/\msun \lsim 10^9$
(see e.g. Ref.~\cite{Ferrarese:2005} for a recent review).

A hint of the existence of IMBHs is provided by the 
detection of bright, X-ray, point sources, called 
ultra-luminous X-ray sources (ULXs), that are 
apparently not associated with active galactic 
nuclei~\cite{Colbert:2002mi,Swartz:2004xt,Dewangan:2005}.  
Although many known X-ray sources are associated 
with neutron stars and black holes, this intepretation fails in 
the case of ULXs.  ULXs would have to emit radiation far 
above the Eddington limit, if $M \lsim 20 M_\odot$, 
and their positions in their host galaxies are not 
compatible with masses $M \gsim 10^6 M_\odot$, 
because dynamical friction would cause these objects to
sink to the centers of their hosts on a timescale 
shorter than a Hubble time (e.g. Ref.~\cite{Miller:2003sc}). 
Accretion by IMBHs has been advocated as a possible explanation~\cite{islamb:2004}.  
Another hint for the existence of IMBHs, 
although not conclusive, comes from stellar kinematics 
in globular clusters~\cite{Frank:1976}; 
the observed relation between the mass and the velocity dispersion  
in selected globular clusters may fall on the extrapolation
of the analogous relation for 
SMBHs~\cite{Gebhardt:2002,vanderMarel:2002,Gerssen:2002}.  

From a theoretical point of view, a population of massive seed 
black holes could help to explain the origin of SMBHs. 
In fact, observations of quasars at 
redshift $z\approx 6$ in the 
Sloan Digital survey ~\cite{Fan:2001ff,barth:2003,Willott:2003xf}
suggest that SMBHs were already in place 
when the Universe was only $\sim 1$ Gyr old, a circumstance 
that can be understood in terms of rapid growth starting 
from massive seeds (see e.g. Ref.~\cite{haiman:2001}).  
Furthermore, the growth of SMBHs through accretion 
and merging of heavy seeds may aid in the understanding some of 
the observed relationships between supermassive black hole 
masses and the properties of their host galaxies and 
halos~\cite{kormendy:1995,Ferrarese:2000se,McLure:2001uf,Gebhardt:2000fk,Tremaine:2002js,Koushiappas:2003zn}.  
Scenarios that seek to explain the properties of the observed 
supermassive black hole population generally result in the 
prediction of a concomitant population of ``wandering'' 
IMBHs throughout massive DM halos and the intergalactic 
medium~\cite{islama:2003,volonteri:2003,Koushiappas:2005qz}.
However, despite their theoretical interest, it is difficult to obtain
conclusive evidence for the existence of IMBHs.  
A viable detection strategy could be the search 
for gravitational waves produced in the mergers of the IMBH 
population~\cite{Thorne:1976,Flanagan:1998a,Flanagan:1998b,islamd:2004,
Matsubayashi:2004,Koushiappas:2005qz}, 
which may become possible with the advent of space-based 
interferometers such as LISA. 

\subsection{IMBHs formation scenarios}

%
%
%
%

We focus here on two scenarios leading to the formation of
black holes at very different mass scales. 
In the first scenario (which we refer to as {\it scenario A}), 
black holes are remnants of the collapse 
of Population III (or ``first'') stars~\cite{Madau:2001}. 
Numerical simulations suggest that 
the first stars may form when primordial 
molecular clouds with $\approx 10^5 \msun$ cool 
by formation and distruction of H$_2$
into cold pockets at the centers of their DM halos, 
with typical densities of order $10^4$~cm$^{-3}$
and temperatures of order a few~$\times 10^{2}$~K 
\cite{Abel:2000tu,Bromm:2001bi},  
and become gravitationally unstable. 

Newtonian simulations suggest that the fate of Pop~III stars is very
different from the case of their metal-enriched, 
comparably less massive counterparts mentioned 
above.  Zero metallicity Pop~III stars with masses in the range 
$M \sim 60 - 140 \msun$ and $M \gsim 260 \msun$ collapse 
directly to black holes, stars with 
$140 \lsim M/\msun \lsim 260 \msun$ are completely 
disrupted due to the pulsation-pair-production instability, 
leaving behind no remnant, and again stars with masses 
$M \gsim 260 \msun$ collapse directly to 
black holes~\cite{Heger:2002by} (see also 
Refs.~\cite{Bond:1984,Fryer:2000my,Larson:1999zs,Schneider:1999us}).  
The evolution timescale of these very massive stars is 
typically of order $t_{*} \sim 1-10$~Myr.  
After this timescale, supernovae begin to explode, releasing 
energy and metals into the surrounding medium.  
In the standard picture of hierarchical structure 
formation, the metal-enriched material will be 
collected at later times at the centers of 
more massive haloes, where new generations of 
stars will form.

Interestingly, if a $\sim 10^2 \msun$ black hole forms 
halos that represent $\sim 3 \sigma$ peaks of the 
smoothed density field, the resulting baryonic mass fraction in these 
objects would be comparable with the mass 
fraction in SMBHs~\cite{Madau:2001}.  
Additionally, such a scenario leads to the 
natural prediction of a population of ``wandering'' 
black holes in the halos of Milky Way-sized galaxies, 
with masses similar to their initial mass scale 
$M \sim 10^2 \msun$, as many of the relatively 
small halos 
($M \sim 10^7 \msun$ for $3 \sigma$ 
fluctuations at $z=18$) 
that host early-forming black holes 
do not merge with the central galaxy, but orbit 
about the periphery of the halo 
\cite{volonteri:2003,islama:2003,Koushiappas:2005qz}.  
We stress that black holes in this scenario 
may not necessarily form at the very centers 
of their initial host dark matter halos at 
high redshift, a circumstance that, as we shall see, may 
have important consequences on the detectability of IMBHs.  

To represent the predictions of this class of 
black hole formation scenario where black holes 
form at $\sim 100 \msun$ from the remnants of the 
first stars, 
we use a model similar to that proposed by 
Madau and Rees~\cite{Madau:2001} and studied in further 
detail by Islam et al.~\cite{islama:2003} and Volonteri et al.~\cite{volonteri:2003}.  
Specifically, at $z=18$, 
we populate halos that constitute 
$3 \sigma$ peaks in the smoothed primordial density 
field with seed black holes of initial mass~$100 \msun$.  
We evolve these halos using an analytic model of halo 
growth that is focused on making many statistical realizations 
of the growth of a Milky Way-sized halo.  After populating 
progenitor halos at high redshift with black holes as 
described above, these processes of halo growth and evolution 
are treated as described in detail in 
\cite{Zentner:2003,Zentner:2005} and 
\cite{Koushiappas:2004,Koushiappas:2005qz}.  
We refer the reader to these references for 
details and tests of the halo evolution models.  
For the purposes of this study, we take the 
mass of the Milky Way halo to be 
$M_{\mathrm{MW}} = 10^{12.1} \ h^{-1}\msun$ 
and perform $200$ statistical realizations 
in order to ascertain the expected range of 
observable IMBHs.

The second scenario that we consider 
({\it scenario B}) is based on the proposal 
of Ref.~\cite{Koushiappas:2003zn}
and it is representative of a class of models
in which black holes originate from 
massive objects formed directly during the 
collapse of primordial gas in early-forming 
halos~\cite{Haehnelt:1993,Loeb:1994,Eisenstein:1995,Haehnelt:1998,Gnedin:2001ey,Bromm:2002hb}.  
In this class of models, the initial black holes 
are massive ($\sim 10^5 \msun$) and the growth of 
SMBH proceeds in such a way that both mergers and 
accretion play an important role.  We use the model 
of Ref.~\cite{Koushiappas:2003zn} to represent the 
predictions of models that start SMBH growth from 
very massive seeds.  The proposal of 
Ref.~\cite{Koushiappas:2003zn} is 
as follows.  During the virialization and collapse
of the first halos, 
gas cools, collapses, and forms pressure-supported disks 
at the centers of halos that are sufficiently massive 
to contain a relatively large amount of molecular 
hydrogen (molecular hydrogen is the primary gas 
coolant in halos in the relevant mass range, see 
\cite{Tegmark:1997} for a review).  In halos that 
are both massive enough that molecular hydrogen 
cooling is efficient and which do not experience 
any major mergers over a dynamical time, 
a protogalactic disk forms and can evolve 
uninterrupted.  An effective viscosity due to 
local gravitational instabilities in the disk 
leads to an effective viscosity that transfers 
mass inward and angular momentum outward 
\cite{Lin:1987} until supernovae in the first 
generation of stars heat the disk and terminate 
this process \cite{Koushiappas:2003zn}.  By the 
time the process terminates 
(of order the lifetimes of Pop~III stars, 
$\sim 1-10$~Myr), a baryonic mass of order 
$\sim 10^5 \msun$ loses its angular momentum 
and is transferred to the center of the halo.  
Such an object may be briefly pressure-supported, 
but it eventually collapses to form a black hole 
\cite{Heger:2002by,Shapiro:1983}.  

The requirements that the early-forming 
host halo be massive enough to form an 
unstable disk and that the halo not experience 
a major merger imprints a typical mass scale 
for halos within which this process occurs 
of order $\sim 10^7 \msun$.  
In this case the characteristic mass of the 
black hole forming in a halo of virial mass 
$M_v$ is given by 
\begin{eqnarray}
\label{eq:mbh}
M_{\rm bh}& = 3.8 \times 10^4 \msun 
\left( \frac{\kappa}{0.5} \right)
\left( \frac{f}{0.03} \right)^{3/2} \nonumber \\
& \left( \frac{M_v}{10^7 \msun} \right)
\left( \frac{1+z}{18} \right)^{3/2}
\left( \frac{t}{10 {\rm Myr}} \right),
\end{eqnarray}
where we have assumed that a fraction $f$ is the fraction 
of the total baryonic mass in the halo that has fallen into 
the disk, $z$ is the redshift of formation, $\kappa$ 
is that fraction of the baryonic mass which loses its 
angular momentum that remains in the remnant black 
hole, and $t$ is the timescale for the evolution of the 
first generation of stars \cite{Koushiappas:2003zn}.  
The distribution of black hole masses is a log-normal 
distribution with a mean given by the characteristic 
mass above and a standard deviation $\sigma_{M_{\rm bh}}=0.9$.  
The spread is determined by the spread in total angular 
momentum exhibited by halos of fixed mass in cosmological 
$N$-body simulations of DM halo formation \cite{Bullock:2000ry}.
Using the prescriptions of the 
Koushiappas et al.~\cite{Koushiappas:2003zn} model, 
we can again populate halos with black holes at high 
redshift and evolve them forward to determine the 
properties of satellite black holes in a statistically 
large sample of Milky Way-like halos at $z=0$.  This 
is precisely what was done in 
Ref.~\cite{Koushiappas:2005qz} in order to study 
the gravity wave background and we refer the reader 
to this work for further details.  As with 
scenario A, we take the 
Milky Way halo to have mass 
$M_{\mathrm{MW}} = 10^{12.1} h^{-1}\msun$ at $z=0$ and 
construct $200$ realizations of wandering black hole 
populations in halos of this mass.  

\subsection{Intermediate-Mass Black Holes in Milky Way-sized Halos}

In the previous two sections, we outlined models for 
the production of IMBHs in the early universe and 
evolution of IMBHs in their host halos in the context of 
the hierarchical CDM model of structure formation.  
Of course, as halos merge to form larger systems 
that eventually grow to the size of the Milky Way, 
black holes merge,  producing supermassive, central 
black holes and perhaps a detectable gravity wave signal.  
These products have been the focus of most previous 
work regarding these models 
\cite{volonteri:2003,Koushiappas:2003zn,Sesana:2004,Koushiappas:2005qz}.  
Consequently, these studies focused much attention on 
the merging of black holes as halos and galaxies merge.  
On the contrary, we are most interested in those pristine  
black holes that are orbiting within the Milky Way 
halo and have not merged with other black holes because 
these unmerged black holes may still reflect 
the properties of the dark matter density 
enhancement in which they formed.

\begin{figure}[t]
\includegraphics[width=3.25in,clip=true]{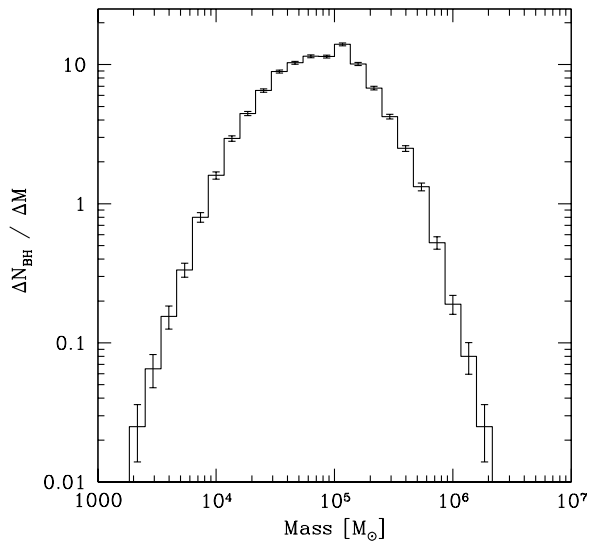}
\caption{\label{figmass} 
%
%
Mass function of unmerged IMBHs in the 
scenario B, for a Milky Way Halo at z=0.  
The distribution is based on an average of 
$200$ Monte Carlo realizations of a halo of 
virial mass $M_v = 10^{12.1} h^{-1}\msun$, 
roughly the size of the halo of the Milky Way.}
\end{figure}

\begin{figure}[t]
\includegraphics[width=3.25in,clip=true]{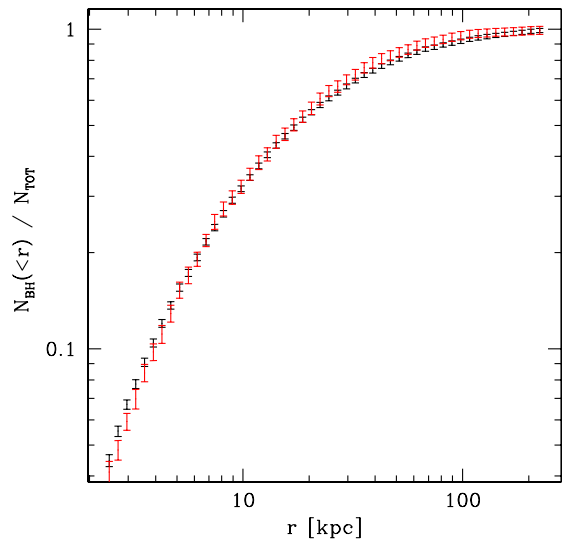}
\caption{
\label{fig:radial} 
Cumulative radial distribution of unmerged IMBHs in the scenario 
A (red) and B (black), for a Milky Way Halo at z=0.  
The mean and error are based on $200$ Monte Carlo 
realizations of IMBH populations in Milky Way-sized 
halos.  Notice that unlike subhalo populations, IMBHs 
do not exhibit a significant anti-bias with respect to 
the DM.  Rather, they are slightly biased toward being 
found near the halo center.
}
\end{figure}

In scenario A, the mass spectrum of unmerged black 
holes is a delta function as described in 
Section~II~A.  The average number of {\it unmerged} black holes 
per Milky Way halo is $N_{\rm bh,A} \simeq 1027 \pm 84$, 
where the errorbar denotes the $1\sigma$ scatter from 
halo-to-halo.  
In scenario B, the total number
of {\it unmerged} black holes per Milky Way halo 
is $N_{\rm bh,B} \simeq 101 \pm 22$.  
We show in Fig.~\ref{fig:radial} the final mass spectrum 
(i.e. at redshift $z=0$) of black holes in scenario B. 
As expected the distribution follow closely the initial 
mass spectrum, with a characteristic mass of order 
$\approx 10^5 \msun$.  The only deviation is that the 
overall distribution is slightly broadened by the fact 
that not all black holes form at the same redshift in 
halos of the same mass (see Eq.~[\ref{eq:mbh}] and 
Refs.~\cite{Koushiappas:2003zn,Koushiappas:2005qz}).  
The radial distribution of unmerged black holes 
is less trivial, and it would be more difficult 
to derive directly from the models of IMBH formation at high redshift.  
The distribution is essentially set by the energy 
and angular momentum distributions of merging objects in 
a {$\Lambda$}CDM cosmology and dynamical friction 
(e.g.~\cite{Zentner:2005}).
Unlike dark matter substructures, 
which are generally absent from the inner 
parts of the host halos, because they tend to lose mass 
via tidal mass loss and heating, 
black holes and the surrounding dark 
matter distribution in the vicinity of the IMBHs 
can survive tidal disruption to very small galactocentric 
distances.  The final, cumulative radial distributions of 
unmerged IMBHs are shown in Fig.~\ref{fig:radial}.  
They are  very similar for scenarios A and B 
(though the normalization is different),
and shows a behavior that scales as 
$dN/dr \sim r^{-3}$ at large scales 
and tends toward a shallower slope on 
scales smaller than the scale radii of 
typical MW-size halos (see the following section).

\section{The Density Enhancement of Dark Matter Around IMBHs}

In each early-forming halo that hosts a seed black hole, 
when the black hole forms the DM distribution about the black 
hole inevitably reacts, adjusting to the new gravitational potential.  
This process has been studied extensively, particularly in 
the context of stellar cusps around massive black holes in clusters of stars 
or at the centers of galaxies 
(see e.g.~\cite{peebles:1972, young:1980, Ipser:1987ru, Quinlan:1995}). 
Gondolo and Silk have applied this argument to the distribution of 
DM at the center of the Galaxy~\cite{Gondolo:1999ef} and 
introduced the term ``spike'' for the consequent enhancement in 
the DM density around the central SMBH, 
in order to avoid confusion with DM ``cusps'' at the centers of halos 
in the {\em cold} dark matter model of structure formation.  
It was subsequently shown that dynamical processes like 
off-center formation of the seed black hole,
or major merger events, may lead to destruction or reduction of the 
spike~\cite{Ullio:2001fb,Merritt:2002vj}.  
However, steeply rising stellar cusps in 
the innermost regions of galaxies suggest that 
such processes were not effective, 
at least in the case of the Milky Way, or that the stellar
cusps were re-generated via star formation~\cite{Milosavljevic:2004te} 
or energy exchange between stars~\cite{Preto:2004kd}.  

Recently, Bertone and Merritt studied the evolution of 
DM spikes including gravitational scattering off stars
and the self-annihilation of DM 
particles~\cite{Bertone:2005hw,Bertone:2005xv}, 
showing that the DM density in spikes is, indeed, 
substantially reduced by these effects, 
but the enhancement of the annihilation 
signal is still significant with respect 
to ordinary DM cusps.

In the present study, we are interested in 
``mini-spikes'' surrounding IMBHs.  Because we 
track the merger history of each individual 
black hole, we can select precisely those black 
holes which never experienced mergers, 
to ensure that major mergers have not destroyed any 
cusp that existed around the original black hole.

Furthermore, the models we explore predict from between 
a few hundred to a few thousand black holes scattered 
throughout the Milky Way halo, and as the Milky Way has 
only $11$ luminous companions within $\sim 300$~kpc 
\cite{mateo:1998}, we expect that the majority of the 
wandering black holes in our models reside in satellite 
halos with no significant stellar component.

This implies that the effects of scattering off of stars 
should not significantly alter the DM distributions around 
the wandering IMBHs.  The mini-spikes around unmerged, 
wandering, IMBHs are thus less sensitive 
to all of the dynamical processes that may have 
affected the spike at the Galactic center.

We proceed now to evaluate the DM enhancements around IMBHs.  As a first step, 
we need to specify the ``initial'' DM profile, that is, the DM distribution prior 
to black hole formation. Let the subscript ``$f$'' denote quantities at 
the time when the IMBH formed. The initial DM profile of the mini-halo, 
before adiabatic growth, can be well approximated with a 
Navarro, Frenk, and White (NFW) profile~\cite{Navarro:1996he}
\begin{equation}
\rho(r)=\rho_0 \left( \frac{r}{r_s} \right)^{-1} \left( 1+\frac{r}{r_s} \right)^{-2} 
\label{eq:nfw}
\end{equation}
The normalization constant $\rho_0$, and the scale radius $r_s$, can
be expressed in terms of the virial mass of the halo at the time when
the IMBH formed $M_{{\rm vir},f}$, and the virial concentration parameter 
$c_{{\rm vir},f}$
\begin{equation}
r_s=\frac{r_{{\rm vir},f}}{c_{{\rm vir},f}} \;\;\;\; , \;\;\;\; 
\rho_0=\frac{M_{{\rm vir},f}}{4\pi r_s^3 f(c_{\rm vir,f})} \;\;.
\end{equation}
We recall that the virial mass is related to the virial radius $r_{{\rm vir},f}$ by
\begin{equation}
\label{eq:mvir}
M_{{\rm vir},f}=\frac{4\pi}{3} \left[\Delta_{\rm vir}(z_f) \rho_m(z_f) \right] r_{{\rm vir},f}^3\;\; ,
\end{equation}
while the function $f(x)$ is, apart from constants, simply 
the volume integral of the NFW profile $f(x) \equiv \ln(1+x)-[x/(1+x)]$.  

In Eq.~(\ref{eq:mvir}), 
$\rho_m(z_f)$ is the mean DM density at the redshift 
of formation $z_f$, 
while $\Delta_{\rm vir}(z_f)$ is the virial overdensity, 
for which we have adopted here the fitting form of 
Bryan and Norman~\cite{Bryan:1997dn}.  At the redshifts of 
interest ($z \gsim 12$) the universe is DM-dominated and 
the expansion rate and growth of perturbations are described 
by the standard relations for an $\Omega_{\rm M} = 1$, 
``standard'' CDM cosmology.  In this case, 
$\Delta_{\rm vir}(z_f) \simeq 18\pi^2 \simeq 178$.  
For each black hole at redshift $z=0$ 
we extract from its merger tree the parameters 
$M_{{\rm vir},f}, \; c_{\rm vir,f}, \; z_f$ 
and use Eq.~(\ref{eq:nfw}) 
to calculate the initial DM profile before the 
formation of the black hole.  
Alternatively, we could have chosen the more 
recent parametrization proposed
by Navarro et al.~\cite{Navarro:04b} (see also
Refs.~\cite{Reed:05,Merritt:05}).  However, this profile 
implies modifications at scales smaller than those we are 
interested in, where the profile is 
anyway modified by the presence of the IMBH.

We assume that the black holes form over
a timescale long enough to guarantee adiabaticity, 
but short compared to the cosmological evolution of 
the host halo (in scenario B, both of these assumptions 
are built into the black hole formation model, see 
Section~II~B as well as our discussion below).  

Adiabaticity requires that the formation time of 
the black hole is much larger than the dynamical 
timescale at a distance $r_h$ from the black hole, 
where $r_h$ is the radius of the sphere of gravitational 
influence of the black hole, 
$r_h \simeq G M_{\rm bh}/\sigma^2$,
and $\sigma$ is the velocity dispersion of 
DM particles at $r_h$.
In practice, we estimate $r_h$ by solving 
the implicit equation 
\begin{equation}
M(<r_h) \equiv \int_{0}^{r_h} \rho(r) r^2 \; {\rm d}r = 2 \; M_{\rm bh}\;\; .
\end{equation}
For a representative case in scenario B, with 
$M_{\rm bh}=10^8 M_{\odot}$ and $M_{{\rm vir},f} = 10^8 M_{\odot}$, 
this gives $r_h/r_s \approx 0.04$.  
In scenario B, the black hole formation time is set 
by the timescale for viscous angular momentum loss and is 
limited by the evolutionary timescale of the first stars and 
the gravitational infall time across the gaseous disk, 
which is of order~Myr (see Ref.~\cite{Koushiappas:2005qz} for 
a detailed discussion of timescales).  The relevant timescale 
for the mass build up of the IMBH is then $t_{ev} \sim 1 - 20$~Myr.  
In scenario A, we follow 
Ref.~\cite{Zhao:2005zr} where the characteristic 
timescale for the growth of the black hole by 
accretion is taken to be of order
$1-20$~Myr for a plausible range of 
accretion efficiencies.  

%
%

The basics of adiabatic growth can be easily 
understood (e.g. Ref.~\cite{Bertone:2004pz}), 
and in most cases the details can be worked 
out by taking into account the approximate 
conservation of adiabatic invariants under a 
certain set of assumptions.  
If one starts from an initially uniform DM 
distribution, the final profile will be a 
mild mini-spike with 
density $\rho_{\rm sp} \propto (r/r_h)^{3/2}$ 
(e.g. see~\cite{Quinlan:1995} and references therein).  
If one starts from a cuspy profile, 
such as the NFW profile of Eq.~(\ref{eq:nfw}), 
the new profile is essentially a power-law, 
\begin{equation}
\rho_{\rm sp}(r)=\rho(r_{\rm sp}) \left(\frac{r}{r_{\rm sp}}\right)^{-\gamma_{\rm sp}} 
\end{equation}
where the radius of the spike is 
$r_{\rm sp} \approx 0.2 r_h$~\cite{Merritt:2003qc},
and $\gamma_{\rm sp}$ is related to the 
initial power-law index $\gamma$ by~\cite{Gondolo:1999ef}
\begin{equation}
\gamma_{\rm sp}=\frac{9-2\gamma}{4-\gamma} \;\; .
\end{equation}
In the case of the profile of Eq.~(\ref{eq:nfw}), 
this reduces to $\gamma_{\rm sp}=7/3$.

The DM annihilation flux in this case diverges at small radii.  
However, the very annihilations that we study here provide an 
upper limit to the DM number (and thus
mass) density.  In absence of other processes affecting 
the distribution of DM, the DM density obeys the equation 
\begin{equation}
\dot{n}_\chi(r,t)=-\sigma v \,\, n^2_\chi(r,t)
\end{equation}
where $\sigma v$ is the annihilation cross-section times
relative velocity (in the non-relativistic limit) and 
$m\chi$ is the DM particle mass. The solution to the evolution equation is 
\begin{equation}
n_\chi(r,t)=\frac{n_\chi(r,t_f)}
{1+n_\chi(r,t_f)  \, \, \sigma v \, \,(t-t_f)}
\end{equation}
which shows that efficent annihilations set an upper limit to the
matter density of order $m_\chi/\sigma v(t-t_f)$.  
We define $r_{\rm lim}$ as the radius where 
\begin{equation}
\rho_{\rm sp}(r_{\rm  lim})=m_\chi/\sigma v \,(t-t_f) \equiv \rho_{\rm lim}
\,\,\, .
\label{eq:rholim}
\end{equation}
We therefore define an inner cut-off at a radius 
\begin{equation}
r_{cut}={\rm Max} \left[ 4 R_{\rm Schw}, r_{\rm lim} \right]
\end{equation}
where $R_{\rm Schw}$ is the Schwarzschild radius of the 
IMBH $R_{\rm Schw} = 2.95 \,{\rm km} \, M_{\rm bh}/\msun$.  
For common values of the mass and cross section of the DM particle, 
$r_{lim} \sim 10^{-3}$~pc so that $r_{cut} = r_{\rm lim}$.  

Is the adiabatic growth of a central mass a good approximation 
in our IMBH formation scenarios?  We have already discussed the 
timescales involved, but the derivation of the inner radius 
$r_{\rm lim}$ provides us with the possibility of checking 
whether the size of the region where matter accretes, 
leading to the formation of a black hole, is actually smaller 
than the characteristic size of the DM spike, $r_{\rm lim}$.  
In scenario A, this is not a problem, because in 
this case the black holes from from Pop.~III stars and the 
spike is produced by the {\it growth} of the black hole, 
thus by processes occuring on scales of order 
$R_{\rm Schw} << r_{\rm lim}$.
In scenario B, the situation is different, 
because the mini-spike is produced by the flow 
of protogalactic material that lost its angular
momentum by viscosity.

Such an object may collapse directly to a 
black hole or it may form a short-lived, 
pressure-supported object 
\cite{Shapiro:1983,Heger:2002by}.
However, in either case, the characteristic size 
of the massive object that forms is likely to 
be much smaller than $r_{\rm lim}$.  
We can make an order-of-magnitude estimate of 
the relative sizes as follows.  
Massive stars are believed to have a polytropic equation of state 
with n=3.  In other words, the equation of state is described by 
\begin{equation}
P(r)=K\rho(r)^\Gamma \, \, , \,\, \Gamma=1+1/n \, , \,
\end{equation} 
and in this case $n=3$ implies $\Gamma = 4/3$, 
as appropriate for a star supported by 
radiation pressure.  It is possible to evaluate 
numerically the properties of a polytropic star in 
hydrostatic equilibrium, for $n=3$ the approximate 
relation $\rho_c=54.2\bar{\rho}$ between the central 
and the average density of the star holds 
(see e.g.~\cite{Chandrasekhar:1967}). 

We infer that the typical scale for the radial 
extent of such an object should be  
\begin{eqnarray}
\nonumber
&R_*=\left( 54.2 \frac{3M_*}{4 \pi \rho_c} \right)^{\frac{1}{3}} \\
\approx&10^{-5} {\rm pc} 
\left( \frac{M}{10^5 \msun} \right)^{\frac{1}{3}}
\left( \frac{\rho_c}{10^{-2} \small {\rm g cm}^{-3}} \right)^{-\frac{1}{3}} \,\,.
\end{eqnarray}
This scale is clearly much smaller than the typical 
size of the spike $r_{cut}$.

\section{Dark matter annihilations in mini-spikes}

Dark matter particles are expected to have a non-negligible 
annihilation cross-section into Standard Model particles, 
in order to be kept in chemical equilibrium in the 
early Universe. It should be a weak or 
weaker-than-weak interaction in order to provide a 
relic density which satisfies cosmological constraints 
(for recent reviews of DM candidates and detection 
techniques see e.g. Refs.~\cite{Bergstrom:2000pn,Bertone:2004pz}). 
Although it is difficult to make definitive statements on the nature 
of the DM particles, it is commonly believed that a mass 
in the range $m_{\chi} \sim 100-1000$~GeV would be a 
reasonable expectation in the most widely-discussed 
DM scenarios (e.g. minimal supersymmetry 
or scenarios with unified extra-dimensions). 
A na{\"{\i}}ve estimate of the annihilation cross section, 
based on the observed relic abundance of DM, 
suggests that $\sigma v \sim 10^{-26}$~cm$^3$s$^{-1}$ . 
This value can be more appropriately used as an 
upper limit to the annihilation cross-section, 
rather than an actual estimate, since processes like 
co-annihilations may significantly affect relic density 
yields (for more details see 
Refs.~\cite{Bergstrom:2000pn,Bertone:2004pz} 
and references therein). 

Instead of 
undertaking a detailed scan of the parameter space for 
different DM candidates, we limit ourselves here to 
estimates of the annihilation fluxes for two 
benchmark models: an optimistic 
model,  with $m_\chi=100$~GeV and 
$\sigma v=3 \times 10^{-26}$~cm$^3$s$^{-1}$, 
leading to large annihilation fluxes; and 
a model with $m_\chi=1$~TeV and 
$\sigma v=10^{-29}$~cm$^3$s$^{-1}$, leading to more
pessimistic predictions.  We note that in both cases, 
the mini-spike profiles reach their maximum values at 
a radii $r_{\rm lim} >> 4R_{\rm Schw}$, thus $r_{\rm lim}$
provides an estimate of the size of the region where 
most of the annihilation radiation originates from.  
The case of annihilations from the DM spike at the center 
of the Galaxy has been extensively studied in the 
literature in terms of neutrino, 
gamma-ray, and synchrotron emission 
~\cite{Gondolo:2000pn,Gondolo:1999ef,Bertone:2001jv,Bertone:2002je,Bertone:2005hw,Bertone:2005xv,Aloisio:2004hy}.  

The flux of gamma-rays from a mini-spike around an IMBH can be 
expressed as 
%
%
\begin{eqnarray}
\nonumber
\Phi(E,D) & = & \frac{1}{2} \frac{\sigma v}{m_\chi^2} \frac{1}{D^2} \frac{{\rm
  d}N}{{\rm d}E} \int_{r_{\rm cut}}^{r_{\rm sp}} \rho^2_{\rm sp}(r) r^2 dr \\
& =  & \frac{{\rm
  d}N}{{\rm d}E} \frac{\rho_{\rm sp}^2}{4\gamma_{\rm sp}-6} \frac{\sigma v}{m_\chi^2} 
\frac{r_{\rm sp}^3}{D^2} \left( \frac{r_{\rm cut}}{r_{\rm sp}} \right)^{-2\gamma_{\rm sp}+3} 
\end{eqnarray}
where we assumed $r_{\rm sp} >> r_{\rm cut}$. 
Inserting typical values of DM and spike parameters we get, 
for the case $\gamma=1$ ($\gamma_{\rm sp} = 7/3$),
\begin{eqnarray}
\nonumber
\Phi (E,D) & = & \Phi_0 \frac{{\rm d}N}{{\rm d}E} 
\left( \frac{\sigma v}{10^{-26} {\rm cm}^3/{\rm s}} \right)
\left( \frac{m_\chi}{100 {\rm GeV}} \right)^{-2} 
\\ 
 & \times & \left( \frac{D}{{\rm kpc}} \right)^{-2} 
\left( \frac{\rho(r_{\rm sp})}{10^2 {\rm GeV}{\rm cm}^{-3}} \right)^{2} 
\nonumber \\
 & \times & 
\left( \frac{r_{\rm sp}}{{\rm pc}} \right)^\frac{14}{3}
\left( \frac{r_{\rm cut}}{10^{-3}{\rm pc}} \right)^{-\frac{5}{3}}\, ,
\label{eq:flux}
\end{eqnarray}
with $\Phi_0 = 9 \times 10^{-10} {\rm cm}^{-2}{\rm s}^{-1}$.  
It is useful here to emphasize the relative 
luminosities of IMBHs in the MW halo.  
In particular, consider the case of 
the relatively more luminous objects of scenario B.  
Using the fiducial values adopted in eq.~(\ref{eq:flux}), 
which are typical of scenario B, one can easily verify 
that {\it the ``luminosity'' of a mini-spike}
(proportional to the volume integral of $\rho_{\rm sp}^2$) 
{\it is of the order of the gamma-ray 
luminosity of the entire Milky Way halo},
a circumstance that has dramatic consequences 
for the prospects of indirect detection, as we 
describe  in the following section.

%
%
\begin{figure}[t]
\includegraphics[width=3.25in,clip=true]{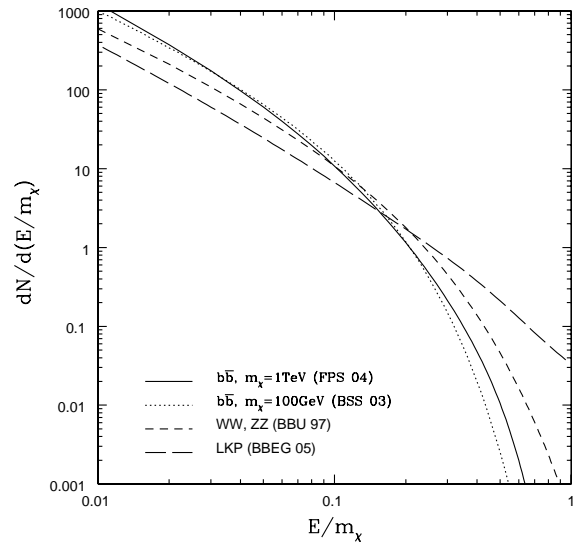}
\caption{\label{fig_dndx} 
Energy spectra of photons per annihilation
for different annihilation channels.  
The {\em solid} and {\em dotted} lines both 
correspond to the $b \bar{b}$ annihilation 
channel, the differences are due to different 
parametrizations of quark fragmentation and 
different DM particle mass scales.  The 
{\em solid} line shows the parameterization of 
Ref.~\cite{Fornengo:2004kj} with $m_{\chi}=1$~TeV, 
while the {\em dotted} line shows that of 
Ref.~\cite{Bertone:2002ms} with $m_{\chi}=100$~GeV.  
The {\em short-dashed} line corresponds to the spectra for 
annihilation through the WW and ZZ channels.  In particular, 
we show the fit from Ref.~\cite{Bergstrom:1997fj}.  
Lastly, the {\em long-dashed} shows the spectrum, 
summed over contributing channels, for annihilation 
of Kaluza-Klein DM from Ref.~\cite{Bergstrom:2004cy}.  
}
\end{figure}
%
To estimate the flux, we need now to specify the gamma-ray 
spectrum per annihilation ${\rm d}N / {\rm d}E$, which depends on 
the nature of the DM particle. In most scenarios, direct annihilation
in two photons is severely suppressed, but a continuum spectrum 
is expected from the decay of secondary neutral pions. 

In Fig.~\ref{fig_dndx} we show the predicted gamma-ray spectra for 
different annihilation channels. For the $b\bar{b}$ channel we
show two different curves, corresponding to different parametrization
of the process of quark fragmentation and subsequent decay of
neutral pions, for 2 different mass scales of the DM particle.
The first (FPS 04) corresponds to the parametrization in 
Ref.~\cite{Fornengo:2004kj}, while the second (BSS 03) 
refers to the spectra presented in Ref.~\cite{Bertone:2002ms}.
The differences for different parametrizations and mass scales appear to be small, 
The third curve (BUB 97) corresponds to an analytic fit for the WW and 
ZZ channels, as discussed in Ref.~\cite{Bergstrom:1997fj} , a channel leading to 
harder spectra with respect to the quark-antiquark channel. These
channels often represent the most important annihilation channels for
neutralino Dark Matter. In the case of Kaluza-Klein DM, other channels
become important.  Following Ref.~\cite{Bergstrom:2004cy} we show in 
Fig.~\ref{fig_dndx} the total spectrum obtained by 
adding the contribution of different channels, 
weighted with the appropriate branching ratios. It is evident from 
the figure that in this case the spectrum is harder than the the quark
or gauge bosons channels, due to contributions from internal bremsstrahlung as
well to decays of quarks and tau leptons. Internal bremsstrahlung is a 
general feature of scenarios where DM particles annihilate into pairs of 
charged fermions, which produces a sharp edge feature in the spectrum, 
dropping abruptly at a photon energy equal to the WIMP mass
~\cite{Beacom:2004pe,Bergstrom:2004cy,Birkedal:2005ep}.

In the next section, we will present our
predictions using the BSS 03 spectrum, 
with the caveat that different annihilation channels may 
lead to slightly different results.  
The predictions for different annihilation channels can be easily
obtained by plugging the appropriate spectrum per annihilation
into Eq.~\ref{eq:flux}.

\section{Results}

In this section, we present our results on the detectability of 
annihilation radiation from the density enhancements about IMBHs. 
The number and properties of the IMBHs population are slightly
different from one realization of Milky Way-sized halos to 
another, as described in Sec. II. To estimate the prospects of
detection of IMBHs in the Milky Way, we thus need to average 
the results over all realizations. 

\begin{figure}[t]
\includegraphics[width=3.25in,clip=true]{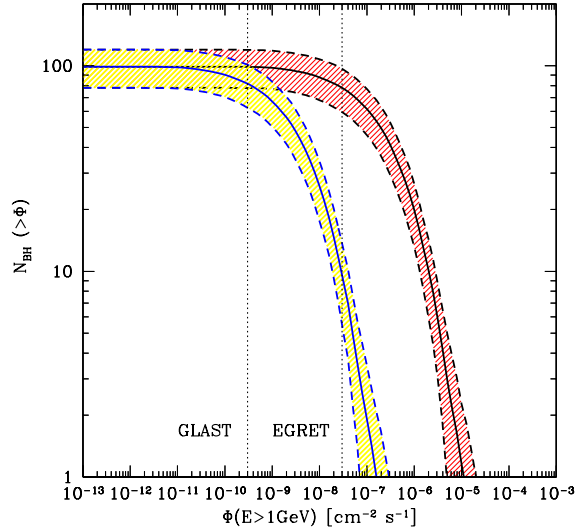}
\caption{\label{fig_detected} 
IMBHs integrated luminosity function,
i.e. number of black holes producing a gamma-ray flux 
larger than a given flux, as a function of the flux, for our
scenario B (i.e. for IMBHs with mass $\sim 10^5 \msun$).
The upper (lower) line corresponds to $m_\chi=100$ GeV, $\sigma v=3\times 10
^{-26}$ cm$^3$ s$^{-1}$ ($m_\chi=1$ TeV, $\sigma v= 10
^{-29}$ cm$^3$ s$^{-1}$). For each curve we also show the 1-$\sigma$
scatter among different realizations of Milky Way-sized host DM halos.  
The figure can be interpreted as the number of IMBHs that can be detected 
from experiments with point source sensitivity $\Phi$ (above 1 GeV), 
as a function of $\Phi$. We show for comparison the 5$\sigma$ point 
source sensitivity above $1$~GeV of EGRET and GLAST (1 year).
}
\end{figure}

In Fig.~\ref{fig_detected}, we show the (average) integrated luminosity 
function of IMBHs in scenario B.  We define the integrated luminosity 
function as the number of black holes producing a gamma-ray 
flux larger than $\Phi$, as a function of $\Phi$.  
The upper (lower) line corresponds to $m_\chi=100$~GeV, 
$\sigma v=3\times 10^{-26}$~cm$^3$s$^{-1}$ 
( $m_\chi=1$~TeV, $\sigma v=10^{-29}$~cm$^3$s$^{-1}$).  
In a practical sense, the plot shows the number of IMBHs
that can be detected with experiments with point 
source sensitivity $\Phi$ above 1~GeV. 
We show for comparison the point source sensitivity 
above 1~GeV for EGRET and GLAST, corresponding roughly 
to the flux for a $5\sigma$ detection of a high-latitude 
point-source in an observation time of 
1~year~\cite{Morselli:2002nw}.  
The dashed region
corresponds to the $1\sigma$ scatter between different 
realizations of Milky Way-sized halos.
This band 
includes the variation in spatial distributions of IMBHs 
from one halo to the next as well as the variation in 
the individual properties of each IMBH in each realization.

Although one would na{\"{\i}}vely expect that the 
fluxes scale with $\sigma v /m_\chi^2$, we note that the 
DM profile itself depends on $m_\chi$ and $\sigma v$, more
precisely on the ratio $\sigma v/m_\chi$ [see Eq.~(\ref{eq:rholim})].  
The maximum density is higher for the pessimistic case, 
and this partially compensates for the decrease in flux due 
to the prefactor $\sigma v /m_\chi^2$.  
It is easy to see this in the case 
$\gamma=1$ from eq.~(\ref{eq:flux}) as, 
by virtue of Eq.~\ref{eq:rholim}, 
$r_{cut} \propto (\sigma v/m)^{-3/7}$,
and the final luminosity of the objects is thus 
proportional to $\sim (\sigma v)^{2/7} m_\chi^{-9/7}$.

The number of detectable sources is very high, 
even in the pessimistic case, 
and either strong constraints on a combination of 
the astrophysics and particle physics of this scenario, 
or an actual detection, should be possible within 
the first year of operation of GLAST, which is 
expected to be launched in 2007.  
Depending on the specific scenario, 
EGRET may have observed some of these IMBH mini-spikes, 
which would still account only for a small fraction 
of the unidentified sources.  

%
%
%

\begin{figure}[t]
\includegraphics[width=3.25in,clip=true]{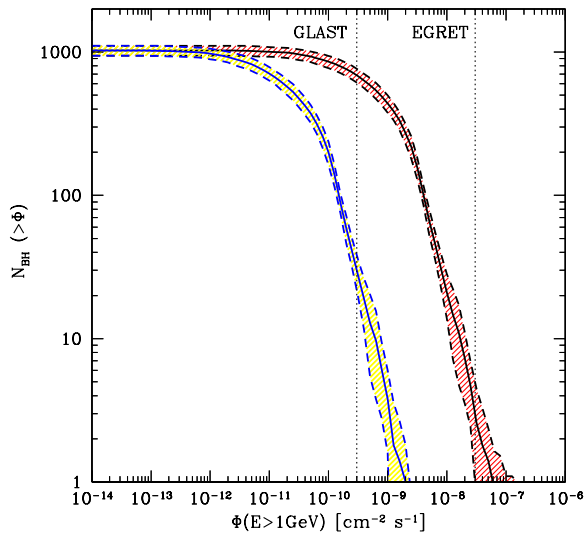}
\caption{ IMBHs integrated luminosity function in scenario A
 (i.e. for IMBHs with mass $\sim 10^2 \msun$).
The upper (lower) line corresponds to $m_\chi=100$ GeV, $\sigma v=3\times 10
^{-26}$ cm$^3$ s$^{-1}$ ( $m_\chi=1$ TeV, $\sigma v= 10
^{-29}$ cm$^3$ s$^{-1}$). For each curve we also show the 
1-$\sigma$ scatter among realizations of Milky Way-sized halos.  
For the sake of comparison, we also show the point source 
sensitivity above $1$~GeV for EGRET and GLAST.}
\label{fig_detectedPEAK}
\end{figure}

We show in Fig.~\ref{fig_detectedPEAK} the 
integrated luminosity function of IMBHs in scenario A, for the
same particle physics models shown in fig.~\ref{fig_detectedPEAK}.  
The lines and error bars all have the same meaning as those 
in Figure~\ref{fig_detected} for scenario B.  
In this case of scenario A, mini-spikes are weaker, 
but the number of black holes is larger by roughly an 
order of magnitude, so that GLAST may still detect 
between a few tens and several hundred sources,
whereas EGRET may have seen only a few or none.  

In figures ~\ref{fig_detected} and ~\ref{fig_detectedPEAK} we have 
assumed that the main annihilation channel is $b\bar{b}$. Although we have 
seen in the previous section that, depending on the nature of the 
DM particle, other channels may dominate and lead to different annihilation
spectra.   We see from these figures that the expected 
uncertainty, $\mathcal O (1)$, would have a small influence on the number of 
objects that GLAST should be able to detect, 
certainly smaller than the uncertainties associated 
with $m_\chi$ and $\sigma v$ and typically smaller
than, or comparable to, the 1-$\sigma$ scatter between different 
Milky Way halo realizations.  

\section{Discussion and Conclusions}

We have studied the detectability of gamma-rays from DM annihilations 
in mini-spikes around IMBHs. The prospects of detection are summarized in
figures~\ref{fig_detected} and~\ref{fig_detectedPEAK}, where we show the number of 
IMBHs that can be detected from experiments with point source sensitivity
$\Phi$ (above 1 GeV), as a function of $\Phi$. We found that the prospects of 
detection with GLAST are so promising that a large number of sources may be
detected within its first year of operation.
With respect to the case of a spike at the 
Galactic center, searching for annihilation radiation from mini-spikes 
has the obvious disadvantage that IMBHs are smaller than the SMBH at 
the Galactic center, and the mini-spikes grow from less dense 
initial profiles.  However, there are also several advantages.  

First of all, 
it is likely that the vast majority of mini-spikes around 
{\em unmerged} IMBHs in the outer Galactic halo are not affected by 
the dynamical processes that tend to destroy the central spike, 
or to decrease significantly the DM density near the black hole.  
For instance, they lack stellar cusps and are rarely 
affected by tidal interactions.  Furthermore, the prospects 
of detectability appear very promising and certainly less 
problematic than, say, annihilations from the Galactic center.  
In fact, the gamma-ray background is strongest in 
the direction of the Galactic center and substantially reduced when 
observations are performed off the disk and in particular at high 
Galactic latitudes.  There is even reason to believe that 
the number density of IMBH may be {\em enhanced} at high 
Galactic latitudes \cite{Zentner:2005b}.  

Moreover, there are several known gamma-ray sources in the direction of the
Galactic center, and the observation of a unique source, even coincident 
with the Galactic center would not necessarily imply a DM annihilation 
origin. On the contrary, the detection of tens or more gamma-ray sources 
with identical spectra, in particular identical cut-offs at the DM
particle mass, and not associated with the Galactic disk or 
other luminous companions of the Milky Way, would
provide a smoking-gun signature of DM annihilations.  

A natural place to search for IMBHs may be the known 
dwarf satellite galaxies about the Milky Way; however, 
the physics that govern the formation of these objects 
is still a topic of much debate and uncertainty, so such 
a search may be subject to the same drawbacks as  searches 
for radiation near the center of the Galaxy.  However, as we 
have already stressed, these IMBH formation scenarios have as a 
virtue that they predict that there may be hundreds of detectable 
objects within the Galactic halo, most of which would not be 
associated with the known population of dwarf satellite galaxies.  
An IMBH population similar to the one in the Milky Way halo 
should be present in Andromeda, given its similarity to the 
Galaxy in terms of size and mass.  
Moreover, the distance to the Andromeda IMBHs would be between 
$\approx 400-1000$~kpc, amounting to a factor of only a few more in
distance than to
the black holes of the outer MW halo. Hence 
the detection of such IMBHs in Andromeda may be possible 
in optimistic scenarios and may serve to demonstrate the 
ubiquity of such phenomena.

As a further implication, we found that the annihilation
luminosity from any Milky Way-like halo may be dominated 
by annihilation around IMBHs in optimistic scenarios.  
As an example, we showed that in scenario B, the total 
luminosity of an individual mini-spike, 
in terms of annihilation radiation, may be 
comparable to the luminosity of the entire host halo.  
Therefore, such optimistic scenarios provide a 
significant ``boost factor'' for the gamma-ray background 
due to DM annihilations in halos at all redshifts 
(e.g., Ref.~\cite{Ullio:2002}, see also Ref.~\cite{ando} for
a comparison between the prospects of indirect detection from the
Galactic center and the gamma-ray background), as well as 
an enhancement of anti-matter fluxes.  
A detailed study of these alternative 
indirect searches requires a full analysis of the redshift 
evolution of IMBHs and spikes in halos of all masses 
in the former case, and of the propagation 
of anti-particles in the Galaxy in the latter, 
and it is thus beyond the scope of this paper.

Interestingly, the prospects of indirect detection in this
scenario do not depend strongly on the particle physics
parameters. In fact, while e.g. in the case of annihilations from the 
Galactic center the annihilation flux is proportional to 
$\sigma v/m_\chi^2$, the flux from mini-spikes is limited by
the plateau in the number density due to DM annihilation itself
For mini-spikes growing from 
$\gamma=1$ profiles, we have shown 
that the annihilation flux is instead 
proportional to $(\sigma v)^{2/7} m_\chi^{-9/7}$. 

Finally, we stress that the detection of these sources 
would shed new light on the origin of IMBHs and SMBHs.  
Mini-spikes appear to be ideal targets for large-field-of-view 
experiments such as GLAST.  Another promising 
experiment with a large field-of-view could be the 
Alpha Magnetic Spectrometer (AMS-02) instrument, 
if preliminary estimates of its sensitivity to 
gamma-rays are confirmed~\cite{brun}. 
Once the positions of the sources are 
determined, Air Cherenkov Telescopes such as 
CANGAROO~\cite{canga}, HESS~\cite{hess}, MAGIC~\cite{magic} 
and VERITAS~\cite{veritas} may provide 
additional information, because of their better 
performance at higher energies
and significantly better angular resolution. The determination
of a common cut-off in the spectra (possible only with ACTs for
DM particles heavier than 300 GeV) will provide an estimate of the
mass of the DM particle, while spectral features, such as 
annihilation lines or sharp edges, may provide important
information on the nature of the DM particle.

%
%
\begin{acknowledgments}

We would like to thank John Beacom, Pierre Brun, Dan Hooper, 
Stelios Kazantzidis, Savvas Koushiappas, David Merritt, 
and Louis Strigari 
for many helpful discussions. GB is supported by the DOE and 
NASA grant NAG 5-10842 at Fermilab.  
ARZ is supported by the Kavli Institute for 
Cosmological Physics at The University of Chicago 
and by the National Science Foundation 
under grant No. NSF PHY 0114422.

\end{acknowledgments}


\end{document}